\documentclass[prl,aps,showpacs,twocolumn]{revtex4-1}

\usepackage{times}
\usepackage{graphicx}
\usepackage{float}
\usepackage{latexsym,amsmath,amssymb,bm,euscript}
\usepackage{color}
\usepackage{epstopdf}
\usepackage[colorlinks=true,linkcolor=blue,citecolor=blue]{hyperref}

\begin{document}

\title{Disorder-driven transition and intermediate phase for $\nu=5/2$ fractional quantum Hall effect}

\author{W. Zhu$^1$ and D. N. Sheng$^2$}
\affiliation{$^1$Theoretical Division, T-4 and CNLS, Los Alamos National Laboratory, Los Alamos, New Mexico 87545, USA}
\affiliation{$^2$Department of Physics and Astronomy, California State University, Northridge, California 91330, USA}

\begin{abstract}
The fractional quantum Hall (FQH) effect  at the  filling number $\nu=5/2$ is a primary candidate for non-Abelian  topological order, while the fate of such a state in the presence of random disorder has not been resolved.
Here, we address this open question by implementing unbiased diagnosis based on numerical exact diagonalization.
We calculate the disorder averaged Hall conductance and the associated statistical distribution of the topological 
invariant Chern number, which unambiguously characterize the disorder-driven collapse of the FQH state.
As the disorder strength increases towards a critical value, a  continuous phase transition is  detected  based on
the disorder configuration averaged wave function fidelity and the  entanglement entropy.
In the strong disorder regime, we identify a composite Fermi liquid (CFL) phase with fluctuating Chern numbers,
in striking contrast to the well-known $\nu=1/3$ case where an Anderson insulator appears.
Interestingly, the lowest Landau level projected local density profile, the wavefunction overlap, and the entanglement entropy
as a function of disorder strength simultaneously  signal an intermediate phase, which may be   relevant  to the recent proposal of  Pfaffian-anti-Pfaffian puddle  state.
\end{abstract}


\maketitle

\textit{Introduction.---}
The fractional quantum Hall (FQH) effect \cite{Tsui1982} is a novel example of topological orders \cite{Wen1990}, providing an ideal testbed for fractional statistics \cite{Laughlin1983} \cite{Moore1991,Greiter1991,Read1999}.
In particular, the non-Abelian quasiparticles following non-commutative
exchange rules,  are expected to form the building-block for topological quantum computation \cite{Kitaev2003,Nayak2008}, thus is of crucial importance.
So far, the even-denominator FQH system at the filling factor $\nu=5/2$ is the most promising candidate for experimental realizations of non-Abelian states
\cite{Willett1987,Pan1999,Choi2008,Pan2008,Radu2008,Dolev2012,Willett2013,Willett2009,Bid2010}.
While this $\nu=5/2$ state was first experimentally identified thirty years ago \cite{Willett1987},
its exact nature remains under  intense theoretical debate.
Among different candidates, the non-Abelian Pfaffian state \cite{Moore1991} as a fully polarized $p_x-ip_y$ paired state of composite fermions \cite{Read2000},
was numerically established as a viable possibility \cite{Morf1998,Rezayi2000,Peterson2008,Hao2009,Storni2010,WZhu2016,Feiguin2008,Jize2011}.
The Pfaffian state breaks particle-hole (PH) symmetry and has a
partner state known as the anti-Pfaffian state \cite{Levin2007,Lee2007} which is also a valid  candidate.
In the presence of an exact PH symmetry, for example by projecting into the first excited Landau level,
the Pfaffian and anti-Pfaffian are exactly degenerate, thus the emergence of one over the other is determined by the  PH symmetry breaking, e.g., through Landau level mixing \cite{Rezayi2011,Zatel2015,Pakrouski2014,Rezayi2017}.
In addition, motivated by the PH-symmetric composite Fermi-liquid (CFL)  at half-filled lowest Landau level \cite{Son2015},
a  non-Abelian PH symmetric Pfaffian (PH-Pfaffian) state  has been proposed as a competing candidate very recently \cite{Feldman2016,JianYang2017,Xie2014}.
Remarkably, these different topological states with different edge excitations can be probed in the thermal Hall measurements.   The thermal Hall conductance of the $\nu=5/2$ state
is found to be $\kappa_{xy} \approx 5/2$ (in units
of temperature times a universal constant $\frac{\pi^2 k^2_B}{3h}$ where  $h$ is the Planck constant and $k_B$ the Boltzmann constant) \cite{Banerjee2018}, which points to the edge structure of PH-Pfaffian rather than Pfaffian or anti-Pfaffian.

The experimental signal of PH-Pfaffian is intriguing and challenge for theoretical understanding.
So far,   existing numerical works \cite{Morf1998,Rezayi2000,Peterson2008,Hao2009,Zatel2015,Pakrouski2014,Rezayi2017}
do not support PH-Pfaffian state in microscopic models with dominant Coulomb interactions.
One possible reason is that the PH-Pfaffian model wavefunction fails to represent a gapped and incompressible phase \cite{Milovanovic2017,Ryan2018,Balram2018}.
Alternatively, the observed $\kappa_{xy} = 5/2$ can be plausibly explained
by disorder-induced mesoscopic puddles made of Pfaffian and anti-Pfaffian states \cite{Feldman2016,Kun2016}.
Compared to  pure systems, there are limited  studies   of the role of random disorder on the $5/2$ state,
which immediately raises some critical questions \cite{Mross2017,Chong2017,Lian2018}: Is PH-Pfaffian state or any other topological phase energetically favorable in a disordered FQH system?
In light of the numerical supports of the Pfaffian (or anti-Pfaffian) in disorder-free systems, another  important question is: what is the fate of the 5/2 FQH state in the presence of disorder?
Generally, when the disorder strength becomes comparable to the strength of interactions between electrons,
the FQH state will eventually be destroyed. A characterization of such  disorder driven transitions is highly desired to compare with experimental observations.
To date, related studies of the disorder FQH systems have only been done at $\nu = 1/3$, where it has been identified a disorder-driven transition from the Laughlin state to an Anderson insulator \cite{DNS2003,DNS2005,ZLiu2016}.
It remains unclear to what extent the above picture will change at $\nu=5/2$.

In this paper, we investigate the disorder-driven transition  for  half-filled first excited Landau level,
based on which, we illustrate a global phase diagram for such a non-Abelian system in the presence of random  disorder.
First of all, we show that the distribution of Hall conductances and the associated topological invariant Chern number  
can be used  to distinguish different quantum phases.   
We identify a disorder-driven critical point separating the FQH state carrying a unique quantized Chern number, from a CFL  that is characterized by a distribution of fluctuating Chern numbers for different disorder configurations.
This is in sharp contrast to the  $\nu=1/3$ FQH state, where the Laughlin state undergoes a transition to an Anderson insulator \cite{DNS2003,DNS2005} with vanishing Chern number at strong disorder side.
This phase transition is also signaled by the variance  of wave function fidelity and the disorder configuration averaged entanglement entropy, both of which support the same critical point for the collapsing of the FQH effect by strong disorder.
In addition, we address the possibility of an intermediate phase in moderate disorder strength, potentially relevant to the disorder
induced  PH-Pfaffian state.
Our  work not only identifies  a novel quantum phase transition between the FQH state and a CFL, but also provides strong evidences  to support the theoretical conjecture  of disorder-stabilized FQH phase  based on numerical simulations of microscopic model for FQH systems\cite{Feldman2016,Kun2016}.

\textit{Model and Method.---}
We consider $N_e$ electrons moving on a torus  under  a perpendicular magnetic field. The torus  is spanned by
${\mathbf L}_1 = L_1 {\mathbf e}_x$ and ${\mathbf L}_2 = L_2 {\mathbf e}_y$, where ${\mathbf e}_x$ and ${\mathbf e}_y$ are  Cartesian unit vectors, and
$L_1$ and $L_2$ are lengths of the two fundamental cycles of the torus.
Required by the magnetic translational invariance, the number of fluxes penetrating a torus is equal to the number of orbitals in one Landau level $N_{s}=L_1L_2/(2\pi \ell^2)$ ($\ell$ is the magnetic length). The total filling fraction is then defined as $\nu=\nu_0+N_e/N_s$ ($\nu_0=2$ for 5/2 FQH systems due to the fully occupied lowest Landau level).
When the magnetic field is strong, we can assume that electrons  in the partially-filled Landau  level are spin-polarized and their dynamics is restricted to the orbitals in the first excited Landau level.
The many-body Hamiltonian can be projected into the first excited Landau level  as
\begin{eqnarray*}
\hat H&=&\sum_{m_{i}=0}^{N_s-1}V^{m_1,m_2}_{m_3,m_4} \hat{a}^{\dagger}_{m_1}\hat{a}^{\dagger}_{m_2}\hat{a}_{m_3}\hat{a}_{m_4}
+ \sum_{m_i=0}^{N_s-1} U^{m_1}_{m_2}\hat a^\dagger_{m_1} \hat a_{m_2}
\end{eqnarray*}
where $a^{\dagger}_{m} (a_{m})$ is the creation (annihilation) operator of an electron in the orbital $m$.
By choosing Landau gauge, the momentum conserved interaction terms can be expressed as
\begin{eqnarray*} \label{torusv}
&&V^{m_1,m_2}_{m_3,m_4}=\frac{1}{4\pi N_s} \delta_{m_1+m_2,m_3+m_4}^{\textrm{mod} N_s}\nonumber\\
&&\,\,\,\, \sum_{q_1,q_2=-\infty}^{+\infty}
\delta_{q_2,m_1-m_4}^{\textrm{mod} N_s}V(\mathbf q)
e^{-\frac{1}{2}|\mathbf q|^2}e^{\textrm{i}\frac{2\pi q_1}{N_s}(m_1-m_3)},
\end{eqnarray*}
where $V({\mathbf q})=\frac{1}{|\mathbf q|}$ represents  the Coulomb interaction and
$\mathbf q=(q_x,q_y)=(\frac{2\pi q_1}{L_1},\frac{2\pi q_2}{L_2})$. The  disorder term is 
\begin{equation*}
U^{m_1}_{m_2}=\frac{1}{2\pi N_s} \sum_{q_1,q_2=-\infty}^{\infty} \delta_{t,m_1-m_2}^{\mod N_s} U(\mathbf q) e^{-\frac{1}{4}|\mathbf q|^2}e^{\textrm{i}\frac{\pi q_1}{N_s}(2m_1-q_2)},
\end{equation*}
where $U(\mathbf q)= \int d\mathbf r e^{i \mathbf q \cdot\mathbf r} U(\mathbf r)$
mimics the random disorder. To study the effects of
correlated potential, we use  the Gaussian correlated
random potential
$ \langle U(\mathbf q)U(\mathbf q')\rangle = \frac{W^2}{2\pi N_s} \delta_{\mathbf q,\mathbf q'} e^{-2q^2\xi^2}$,
where $\xi$ is the correlation length.

We obtain the ground state $\{ |\Phi_k\rangle \}$ of $\hat H$ using exact diagonalization (ED) algorithm.
Due to the lack of translational symmetry in the presence of disorder,
the system sizes accessible by ED  are limited to $N_e\leq12$ by the current computational capability.
In our extensive tests, $N_e\leq8$ systems suffer from very strong finite size effect, so we will focus on the $N_e=10,12$ below. 
We averaged up to $2000$ and $500$ samples for $N_e = 10$ and  $N_e=12$, respectively, which gives quantitatively reliable results.

\textit{Statistics of Chern number.---}
Identifying topological invariant is  crucial for characterizing  the underlying physics of topological ordered states. Conventionally,
FQH states are characterized by the Hall conductance and the associated Chern number \cite{Thouless1982,QNiu1985,Avron2003},
which determines the intrinsic topology of wave function \cite{Kohmoto1985} and the corresponding
gapless edge excitations at a  system  boundary \cite{Hatsugai1993}.
In the presence of disorder, the Hall conductance also offers an unambiguous criterion to distinguish
the insulating state from quantum Hall states in an interacting system \cite{Arovas1988, DNS2003, DNS2005}. To be specific,
under twisted boundary condition the wavefunction becomes 
\begin{equation*}
|\Psi_k \rangle= \exp \left [-i \sum _{i=1}^{N_e} \left (\frac {\theta_1}{L_1} x_i
+ \frac {\theta_2}{L_2}y_i \right ) \right ] |\Phi_k\rangle,
\end{equation*}
and the boundary phase averaged Hall conductance
is $\sigma_H(k)= C_ke^2/h$,
where $C_k$ for the state is defined as \cite{DNS2003}
\begin{eqnarray*}
\label{integral}
C_k ={i\over 4\pi} \oint_{\Gamma} d {\bf \theta} \cdot
\left [ \langle { \Psi_k |{\partial \Psi_k
 \over
\partial {\bf \theta}}\rangle -
\langle {\partial \Psi_k \over \partial {\bf \theta}}|
\Psi_k}
\rangle \right ].
\end{eqnarray*}
Here, the closed path integral is carried out along the boundary
$\Gamma$ of the boundary parameter  space (the magnetic Brillouin zone)   
$0 \leq \theta_1, \theta_2 \leq 2\pi$.
$C_k$ is equivalent to the Berry phase (in units of $2\pi$) accumulated
when the boundary conditions evolve along the closed path $\Gamma$.

\begin{figure}[t]
 \includegraphics[width=0.95\linewidth]{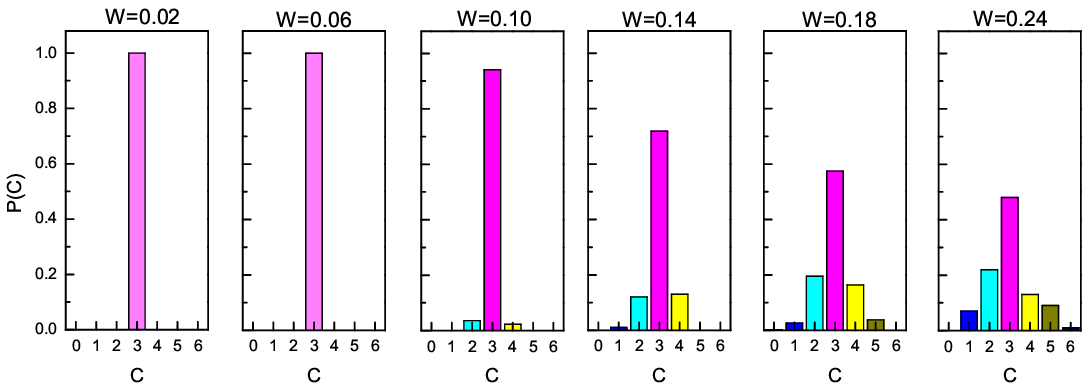}
 \includegraphics[width=0.95\linewidth]{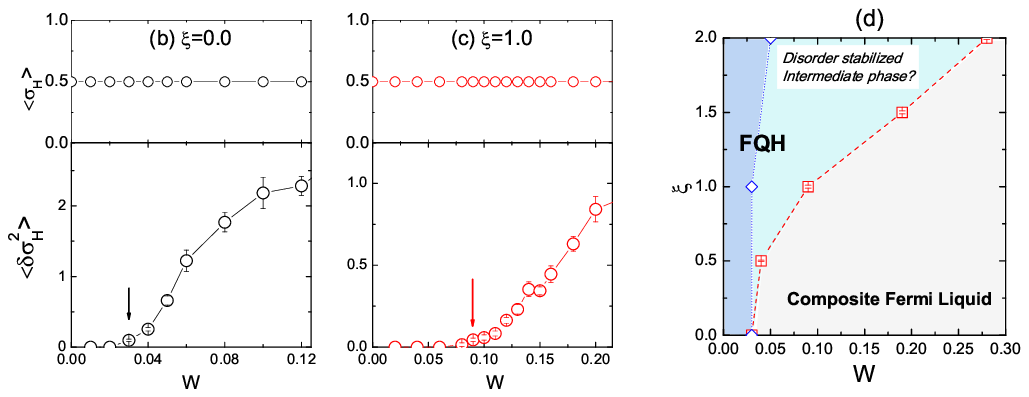}
 \caption{(a) Probability distribution $P(C)$ of total Chern number $C$ for various disorder strength $W$.
 Here we set $\xi=1.0$ for a system with  $N_e=10$ electrons.
The Hall conductance $\sigma_H$ and its fluctuation $(\delta \sigma)^2_H$ versus disorder strength $W$ for (b) $\xi=0.0$ and (c) $\xi=1.0$.
The error bar shows the standard error bar  in the disorder averaged value.
(c) The global phase diagram for $\nu=5/2$ illustrates the FQH phase, the  disorder induced CFL, and 
the possible intermediate phase as labeled in light blue.
 }\label{fig:chern}
\end{figure}

Let us start by discussing the salient features of the Chern number statistics for different disorder strength.
We tune the aspect ratio $L_1/L_2$ to find energy spectrum with six fold near degeneracy separated
from other excited states, which characterizes the particle-hole symmetrized Pfaffian state \cite{Rezayi2000}.  
Taking into account that the lowest six states should become degenerate in the thermodynamic limit,
we introduce probability $P(C)$ of the total Chern number distribution, which describes the probability that total Chern number
of the lowest $N_g=6$ near degenerating  states is $C$ in our sampled configurations.
For a weak disorder strength (Fig. \ref{fig:chern}(a)), $P(C)$ takes unity for $C=3$ and zero for $C\neq3$
(i.e., the lowest six states have $C=3$ for all the disorder configurations), thus that  each nearly degenerated ground state carries a Hall conductance of $\sigma_H=e^2/2h$, which manifests the $\nu=5/2$ FQH state on a torus.

In strong disorder regime, disorder tends to change the Chern number of each state, and  redistributes the probabilities of 
different Chern numbers.
As shown in Fig. \ref{fig:chern}(a), when $W>0.1$, $P(C)$ becomes nonzero for $C\neq3$,
with nearly equal probabilities for  Chern numbers larger or smaller than 3 to appear in different  disorder configurations.   
For example, at $W=0.1$, $P(C=3)$ is reduced to $0.95$ while $P(C=2)\approx 0.025$.
Upon increasing disorder strength,
$P(C=3)$ monotonically decreases  and the distribution of $P(C)$ becomes broader.
The coexistence of different Chern numbers characterizes the delocalization of quasiparticle excitations.
In particular, even though $P(C)$ has a broad distribution instead of a single nonzero value,
we identify the averaged Chern number remains approximately quantized to $\langle C \rangle \approx 3$,
for example, $\langle C \rangle \approx 2.98$ at $W=0.24$ (see Fig. \ref{fig:chern}(a)).
This observation demonstrates each ground state still carries nonzero averaged Hall conductance  in strong disorder regime,
which is consistent with  a CFL rather than an Anderson insulator.  \cite{note1}
A plausible understanding comes from the fact that, various FQH $\nu=5/2$ states such as Pfaffian and anti-Pfaffian,
can be interpreted as pairing states built on a half-filled \textit{CFL} \cite{Halperin1993,Son2015} with different underlying pairing symmetries \cite{Read2000}.
While the transition follows the destruction of the pairing mechanism by disorder, disorder cannot localize  composite fermions at half filling,
since the backscattering and localization are
suppressed due to the intrinsic $\pi-$Berry phase \cite{Scott2016,Scott2017,Haldane2004} \cite{Ando1998,Nomura2007}.
As a comparison, in the case of $\nu=1/3$ FQH, strong disorder destroys the quantization of the
Chern number and leads to $\langle C\rangle\approx 0$, which suggests a topologically trivial Anderson insulator in disorder dominating regime \cite{DNS2003,DNS2005}. Notice that in both cases the Landau level mixing effect is not 
considered,  which may eventually destroy the CFL phase  when the disorder strength exceeds the gap between different
Landau levels.

To quantify the evolution of Chern number statistics with respect to disorder strength,
we demonstrate the fluctuation of the Hall conductance $\delta \sigma_H^2$
as a function of disorder strength $W$ in Fig. \ref{fig:chern}(b-c).
In the weak disorder regime, we observe that Hall conductance carried by each ground state is always quantized to $\sigma_H=e^2/2h$
and its fluctuation is vanishing small $\delta \sigma_H^2\approx 0$.
In strong disorder regime, despite $\langle \sigma_H\rangle$ is quantized,
the broad Chern number distribution leads to a finite fluctuation of the Hall conductance $\delta \sigma_H^2\neq 0$.
We can identify  a critical disorder strength $W_c$  separating a FQH state with zero fluctuation from a critical state with finite fluctuations as marked by arrows in Fig. \ref{fig:chern}(b-c).
The above picture holds for all correlation length  $\xi$ and system sizes we tested.

\begin{figure}[t]
	\includegraphics[width=0.99\linewidth]{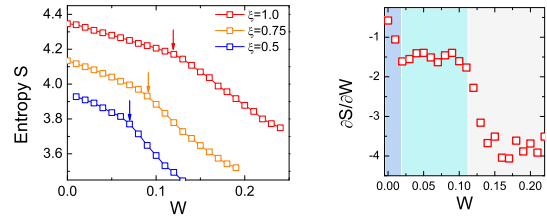}
	\caption{(a) Entanglement entropy $S$ versus disorder strength $W$ of $N_e=10$ electrons for various  correlation length $\xi$. The data  for different $\xi$ is shifted in the vertical direction for clarity.
		(b) Derivative of entropy with respect to the disorder $\partial S/\partial W$ for  $\xi=1.0$.
	}\label{fig:entropy}
\end{figure}

\textit{Entanglement Entropy.---}
Topological phases are characterized by the long-range quantum entanglement patterns \cite{Amico2008,Eisert2010,Laflorencie2016}.
As a novel application, it is found that the entanglement entropy is sensitive to the quantum criticality, in both clean systems \cite{Haque2007,Lauchli2010} and disordered Abelian FQH systems \cite{ZLiu2016,ZLiu2017}.
Fig. \ref{fig:entropy} shows the evolution of entropy by increasing disorder strength at $\nu=5/2$. \cite{note2}
We find that the entropy $S$ monotonically decreases with the increase of $W$. Importantly,
a kink develops near the critical strength $W_c$ (indicated by arrows in Fig. \ref{fig:entropy}(a)), where the slope of entropy shows discontinuity (Fig. \ref{fig:entropy}(b)).
This sudden change of $\partial S/\partial W$ shows a consistent  signature of the expected quantum phase transition. Moreover,
we also identify a trend of the increasing
of the critical $W_c$ for  larger  value of $\xi$.
Importantly,  the entanglement measurements
give consistent identifications of the quantum critical point $W_c$, compared to that identified by Chern number statistics (Fig. \ref{fig:chern}).

\begin{figure}[b]
	\includegraphics[width=0.5\textwidth]{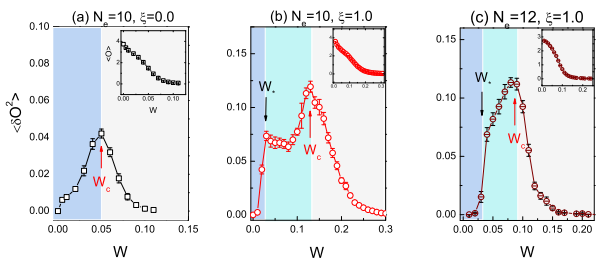}
	\caption{(a-c) Fluctuation of wave function fidelity as a function of disorder strength. Inset: Wave function fidelity versus $W$.
	}\label{fig:overlap}
\end{figure}

\textit{Implications for an intermediate phase.---}
The evolution of Hall conductance and its fluctuation unambiguously pin down the phase transition between the $5/2$ FQH state and CFL state.
However, it is incapable to distinguish the precise nature of different  FQH states, because all candidates, including Pfaffian, anti-Pfaffian or PH-Pfaffian state, carry the same Hall conductance.
Next we further explore the phase transition at the wave function level. 
First, we define  the wave function overlap matrix:
$\langle O\rangle_{ij} = \langle \Phi_i(W)|\Phi^{\mathrm{Pf}}_j(W=0) \rangle$, between the lowest six states  for disordered system with the Pfaffian states,
and   the total overlap $\langle O\rangle$ (fidelity) as the summation of  eigenvalues of the overlap matrix, where $\langle .. \rangle$ indicates the average over the disorder configurations.
In Fig. \ref{fig:overlap}(inset), we show that the wave function fidelity monotonically decreases with the  increase of the disorder,
which does not show a clear signature  of the possible quantum phase transition between different  FQH states.
However, we notice that  the fluctuation of wave function fidelity $\langle (\delta O)^2\rangle$ is sensitive to the phase transition.
This is because in the pure system, the wavefunction is characterized by Pfaffian (anti-Pfaffian) wavefuction, which is a product of Laughlin state for bosonic $\nu=1/2$ and a $p_x\pm ip_y$ wavefunction for composite Fermions.
Physically, the fluctuation of wave function fidelity $\langle \delta O^2\rangle$ can detect the phase fluctuations of wavefunction deviating from the $p_x\pm i p_y$ form.
As shown in Fig. \ref{fig:overlap}, the quantity $\langle \delta O^2\rangle$ demonstrates  a peak structure around the critical point $W_c$.
$\langle \delta O^2\rangle$ reaches a maximum value indicating that the probability distribution of the wave function overlap has largest fluctuations,
indicating critical behavior near the phase transition point.
Interestingly, in addition to the peak around $W_c$, we identify a sudden jump in $\langle \delta O^2\rangle$ at $W_*$ before the FQH to CFL transition with   $W_*<W_c$. This sudden jump takes place at finite $W_*\neq0$ for correlated disorder $\xi\neq0$ (Fig. \ref{fig:overlap}(b-c)), while at a vanishing small value $W_*\approx 0$ for uncorrelated disorder $\xi=0$ (Fig. \ref{fig:overlap}(a)).
This observation signals  an intermediate phase stabilized by correlated disorder.

\begin{figure}[t]
	\includegraphics[width=1.0\linewidth]{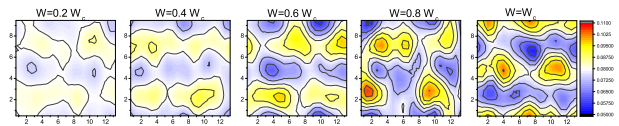}
	\caption{The projected electron density $\rho(\mathbf r)$ for various disorder strength for  $\xi=1.0$ for  systems with  $N_e=10$.
	}\label{fig:puddle}
\end{figure}

To inspect the effect of disorder in real space,   we show  the projected electron density $\rho(\mathbf r)$ in Fig. \ref{fig:puddle},
which is the equivalent electron density describing the
spatial distribution of the guiding center \cite{Rezayi1999,Rezayi2000,DNS2002}.
The many-body density of states is qualitatively distinguishable from the pure limit:
Density modulation is pronounced in spatial space and forms puddle structures starting from  $W\gtrsim W_*$. 
The puddle-like structure provides a consistent microscopic condition for the forming of  an intermediate phase stabilized by correlated disorder.

Based on the appearance of additional critical strength $W_*$ in the variance of the wave function fidelity and puddle formation approximately within $W\gtrsim W_*$,
we identify a possible intermediate phase stabilized by correlated disorder. At quantitative level, nonzero correlation length pushes the critical $W_c$ to larger value (Fig. \ref{fig:chern}(d))
leaving wider  region for the intermediate phase,
which again shows that the intermediate phase is favored by correlated disorder.
All of these are consistent with the puddle picture for the disorder stabilized PH-Pfaffian state \cite{Feldman2016,Mross2017,Chong2017,Lian2018}.
Accordingly, we label an intermediate FQH phase in phase diagram (Fig. \ref{fig:chern}(d)).

\textit{Summary and Discussion.---}
We have presented a systematic  numerical study of correlated disorder driven quantum phase transitions  for $\nu=5/2$ fractional quantum Hall effect.
First of all, the distribution of topological Chern numbers and corresponding Hall conductance fluctuations
are capable of directly probing the collapse of the fractional quantum Hall state, which also determines  the quantum critical points for random  disorder  with  different correlation lengths.
Second, the phase transition is also signaled by
the wave function fidelity and entanglement entropy.  The critical disorder strength  obtained from different methods
is consistent  with each other,  validating the reliability of our numerical results.
Third, in strong disorder regime, we identify a composite Fermi liquid as the ground state, rather than an Anderson insulator as realized at filling number $\nu=1/3$,
demonstrating rich physics for strongly correlated  disorder systems.
Last but not least, our results imply a possible intermediate phase stabilized by correlated disorder potentials,
as evidenced by fluctuations of wave function fidelity and the puddle-like structures in projected density of states.
These results provide the essential step towards understanding the  nature of the disorder stabilized 
5/2 quantum Hall state  in the half-filled first excited Landau level  from a microscopic point of view.
Furthermore, our work indeed opens up several directions for further exploration. 
For example, to connect with the previous studies on network models \cite{Chong2017,Mross2017},
it is important  to identify the neutral chiral modes on the domain walls between randomly distributed puddles. 
In addition, diagnosis of quantum fluctuations via various quantities shown here provides a practical way to study quantum
criticality for general  disordered interacting fractionalized topological   systems.

\textit{Acknowledgements.---}
We thank F. D. M. Haldane,  
 Bo Yang, Zhao Liu, Chong Wang, Yin-chen He, and Jie Wang for simulating discussions.
W.Z. is supported by the U.S. Department of Energy (DOE), 
 through LDRD program at Los Alamos National Laboratory.
D.N.S. is supported by the  U.S. DOE, Office of Basic Energy Sciences under Grant No. DE-FG02-06ER46305.
D.N.S. also acknowledges the travel support by the Princeton MRSEC 
through the National Science Foundation Grant  MRSEC DMR-1420541.


\begin{thebibliography}{99}
\bibitem{note1}Further tests with adding PH symmetry breaking three body interactions confirm that the averaged Chern number remains unchanged even when PH symmetry is broken.
\bibitem{note2}Since the FQH phases host the degenerating states in the thermodynamic limit, we consider a ground state manifold containing the lowest six eigenstates to compute density matrix $\rho$ and related entanglement entropy.
\end{thebibliography}
\end{document}